\documentclass[aps,graphicx,showpacs,amsmath]{revtex4}
\usepackage{amssymb}
\usepackage{graphicx}

\begin{document}


\title{Self-error-rejecting photonic qubit transmission in polarization-spatial
modes with linear optical elements\footnote{published in Sci. China-Phys. Mech. Astron. \textbf{60}, 120312 (2017)}}
\author{Yu-Xiao Jiang$^{1}$, Peng-Liang Guo$^{1}$, Cheng-Yan
Gao$^{1}$,  Hai-Bo Wang$^{1}$, Faris Alzahrani$^{2}$, Aatef
Hobiny$^{2}$,  Fu-Guo Deng$^{1,2,}$\footnote{Corresponding author:
fgdeng@bnu.edu.cn}}

\address{$^{1}$Department of Physics, Applied Optics Beijing Area Major Laboratory,
Beijing Normal University, Beijing 100875, China\\
$^{2}$NAAM-Research Group, Department of Mathematics, Faculty of
Science, King Abdulaziz University, P.O. Box 80203, Jeddah 21589,
Saudi Arabia}

\date{\today }

\begin{abstract}
We present an original self-error-rejecting photonic qubit
transmission scheme for both the polarization and spatial states of
photon systems transmitted over collective noise channels. In our
scheme, we use simple linear-optical elements, including half-wave
plates, 50:50 beam splitters, and polarization beam splitters, to
convert spatial-polarization modes into different time bins. By
using postselection in different time bins, the success probability
of obtaining the uncorrupted states approaches 1/4 for single-photon
transmission, which is not influenced by the coefficients of noisy
channels. Our self-error-rejecting transmission scheme can be
generalized to hyperentangled N-photon systems and is useful in
practical high-capacity quantum communications with photon systems
in two degrees of freedom.
\end{abstract}

 \maketitle

\section{Introduction}

A photon can serve as an ideal qubit because of its maneuverability;
thus, it is regarded as an essential quantum system for realizing
quantum communication, including quantum teleportation
\cite{QT1,QT2}, quantum dense coding \cite{DC1,DC2}, quantum key
distribution \cite{QKD0,QKD1,QKD2,QKD3}, quantum secret sharing
\cite{QSS1}, and quantum secure direct communication
\cite{QSDC1,QSDC2,QSDC3,QSDC4,QSDC5,QSDC6,QSDCreview}. However,
during transmission over an optical-fiber or free-space channel,
photons are inevitably influenced by the thermal fluctuations,
vibrations, and imperfections of the fiber, i.e., noise in the
quantum channel \cite{r2,r3,r5}, leading to phase shifts in spatial
modes and changes in polarization states. Error-rejecting qubit
transmission is an effective method to resist the influence of
collective noise \cite{r5}. In 2005, Kalamidas \cite{r2} proposed
two linear-optical single-photon schemes to reject and correct
arbitrary qubit errors without additional qubits. In the same year,
Yamamoto \emph{et al.} \cite{r3} developed a single-photon
error-rejecting scheme with an ancillary qubit in fixed
polarization. In this scheme, an ancillary particle and a photon in
an arbitrary quantum state pass through the same apparatus and are
affected by the noise channel identically. Detection of the outports
of two photons could exclude errors, thereby probabilistically
giving the original quantum state, the fidelity of which is 1. The
proposed scheme presented a low success probability, which is
relevant to the parameters of noise. Application of photon detectors
and a two-qubit entanglement gate yielded a success probability of
only 1/8. In 2007, Li \emph{et al.} \cite{r5} presented a very
useful scheme for passively error-rejecting qubit transmission using
linear optics and postselection in different time bins without
ancillary qubits. In their scheme, qubits were encoded in time bins,
and uncorrupted photons arrived at definite time slots; this scheme
promised a 50\% success probability in the case of linear-optical
elements only. If time delayers were applied, the success
probability could approach 100\%, regardless of the coefficients of
noise.

Hyperentanglement, which is defined as the simultaneous entanglement
in two or more degrees of freedom (DOF) of a photon system
\cite{dengreview,prep1,prep2,prep3,prep4,prep5,prep6}, has attracted
great attention recently as it can improve the capacity of quantum
communication \cite{HBSA1,HBSA2,HBSA3,HBSA4,HBSA6,HBSA7}  and speed
up the quantum computation
\cite{hypercnot1,hypercnot2,hypercnot3,hypercnot4}. For example, it
can be used to complete the task of teleporting an unknown quantum
state in more than one DOF and double the channel capacity of
quantum communication via the complete hyperentangled Bell-state
analysis \cite{HBSA1,HBSA2,HBSA3,HBSA4,HBSA6,HBSA7}.
Hyperentanglement can also be used to beat the channel capacity
limit for linear photonic superdense coding \cite{HESC}, achieve the
deterministic entanglement purification
\cite{DEPP1,DEPP2,DEPP3,DEPP4,EPPblind}, and construct
high-efficiency quantum repeaters \cite{wangrepeater}. In practical
applications of hyperentanglement in high-capacity quantum
communication, hyperentangled photon systems suffer from channel
noise, as well as photonic quantum states with one DOF. Two
approaches can depress the influence of noise on hyperentangled
quantum systems. The first approach is the passive method, including
hyperentanglement purification
\cite{HEPP1,HEPP2,HEPPWangGY,DuFFHEPP,HEPP5} and hyperentanglement
concentration
\cite{HECP1,HECP2,HyperC3,HyperC4,HyperC5,HECP6,HyperC7,HyperC8,HyperC9,HECPadddu2},
through which the parties in long-distance quantum communication can
distill some high-fidelity hyperentangled photon systems from
nonlocal less-hyperentanglement systems polluted by channel noise.
The second approach is the active method, including error-rejecting
qubit transmission and error-correcting code. This method uses
special DOFs of photon systems or decoherence-free subspace to
protect photonic systems when they are transmitted over a collective
noise channel. These two approaches are necessary to overcome the
influence of channel noise in practical long-distance quantum
communication. To date, no works on error-rejecting qubit
transmissions for hyperentangled photon systems with linear-optical
elements have been published.

In this paper, we first present an original scheme for
self-error-rejecting single-qubit transmission for photon systems in
both the spatial and polarization modes using linear-optical
elements only. The system consists of an unbalanced polarization
interferometer based on a polarizing beam splitter, a 50:50 beam
splitter, and a half-wave plate. We also apply several additional
BSs to overcome the phase shift in the spatial-mode DOF. For a
single-photon transmission, the success probability of receiving
uncorrupted photons is  25\%. We apply this method to the
transmission of two hyperentangled photons and entangled n-photons
and demonstrate its applicability to future high-capacity quantum
communication.

\begin{figure*}[htb]                    
\centering
\includegraphics[width=14 cm]{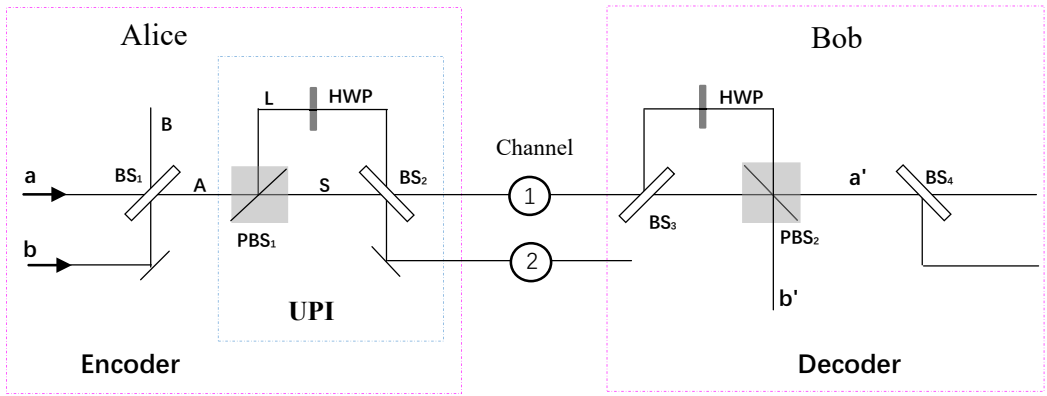}
\caption{Schematic diagram of the error-correcting distribution of a
single-photon in polarization-spatial modes. $a$ and $b$ represent
the two spatial modes of the photon. $A$ and $B$ represent two
different paths for the photon. PBS$_i$ ($i=1,2$) represents a
polarizing beam splitter, which transmits photons with a horizontal
polarization and reflects photons with a vertical polarization.
BS$_{j}$ ($j=1,2,3,4$) is a 50:50 beam splitter. HWP denotes a
half-wave plate which is used to complete the transformations
between the polarizations $|V\rangle$ and $|H\rangle$, i.e.,
$|V\rangle \rightleftarrows |H\rangle$.  UPI represents an
unbalanced polarizing interferometer.
}\label{figure1}
\end{figure*}

\section{Error-rejecting single-photon transmission in polarization-spatial modes}

We suppose that the transmitted photon $P$ from Alice is in an
arbitrary polarization-spatial  state
\begin{eqnarray}            
|\Psi\rangle_P =
(\alpha'|a\rangle+\beta'|b\rangle)\otimes(\alpha|H\rangle+\beta|V\rangle).
\end{eqnarray}
Here $|\alpha|^2+|\beta|^2=|\alpha'|^2+|\beta'|^2=1$. $\vert
H\rangle$ and $\vert V\rangle$ represent the horizontal and vertical
polarizations of photons, respectively. $|a\rangle$ and $|b\rangle$
denote the different spatial modes of the  photon $P$.  The
principle of our scheme for the error-rejecting single-photon
transmission in polarization-spatial modes is shown in
Fig.\ref{figure1}. The photon $P$ in the spatial mode $|b\rangle$
apparently passes through a long path, causing a time delay in the
order of a few nanoseconds. We represent the time difference between
the two spatial modes  $|a\rangle$ and $|b\rangle$ by $\delta t$.
BS$_{1}$ redistributes the paths a and b to two new paths A and B,
and the state of the photon $P$ after it is divided into A and B can
be described as follows:
\begin{eqnarray}            
|\Psi\rangle_P&=&(\alpha'|a\rangle+\beta'|b\rangle)\otimes(\alpha|H\rangle+\beta|V\rangle)
\nonumber\\
&\xrightarrow[]{BS_1}&\frac{1}{\sqrt{2}}(\alpha'|a\rangle_{0}
+i\beta'|b\rangle_{\delta t})(\alpha|H\rangle+\beta|V\rangle)_{A}
+\frac{1}{\sqrt{2}}(\beta'|b\rangle_{\delta t}
+i\alpha'|a\rangle_{0})(\alpha|H\rangle+\beta|V\rangle)_{B}\nonumber\\
&\equiv& |\psi\rangle _{A} + |\psi\rangle _{B}.
\end{eqnarray}
Here
\begin{eqnarray}            
 |\psi\rangle _{A}&=&\frac{1}{\sqrt{2}}(\alpha'|a\rangle_{0}
+i\beta'|b\rangle_{\delta t})(\alpha|H\rangle+\beta|V\rangle)_{A},
\nonumber\\
|\psi\rangle _{B}&=&\frac{1}{\sqrt{2}}(\beta'|b\rangle_{\delta t}
+i\alpha'|a\rangle_{0})(\alpha|H\rangle+\beta|V\rangle)_{B}.
\end{eqnarray}
In either path A or B, states $|a\rangle$ and $|b\rangle$ are
distinguished by a time delay. In addition, the quantum states in
these two paths can be transformed into each other by a unitary
operation; thus, we only consider the outcome from path A in the
following discussion. The outcome from path B is similar to that
from path A.

The photon in path A passes through an UPI.  PBS$_{1}$ in the
encoding system transmits the horizontal polarization mode
$|H\rangle$ and reflects the vertical polarization mode $|V\rangle$,
thus allowing states $|H\rangle$ and $|V\rangle$ to propagate
through the  short path $S$ and the long path $L$, respectively.
This length difference also results in a time delay in the order of
a few nanoseconds. The HWP  transforms $|V\rangle$ to $|H\rangle$.
The transformation of the state for the photon P in path A before it
enters the noisy channel could be described as follows:
\begin{eqnarray}            
|\psi\rangle _{A}&=&\frac{1}{\sqrt{2}}(\alpha'|a\rangle_{0}
+i\beta'|b\rangle_{\delta t})(\alpha|H\rangle+\beta|V\rangle)
\nonumber\\
&\xrightarrow[HWP]{PBS_1}&\frac{1}{\sqrt{2}}(\alpha'|a\rangle_{0}
+i\beta'|b\rangle_{\delta t})(\alpha|H\rangle _{S}+\beta|H\rangle
_{L})
\nonumber\\
&\xrightarrow[]{BS_2}&\frac{1}{2}(\alpha'|a\rangle_{0}
+i\beta'|b\rangle_{\delta t})[(\alpha|H\rangle _{S} +i\beta|H\rangle
_{L})_{1}+(i\alpha|H\rangle_{S}+\beta|H\rangle_{L})_{2}]\nonumber\\
&\equiv& |\psi\rangle _{1} + |\psi\rangle _{2}.
\end{eqnarray}
Here
\begin{eqnarray}            
|\psi\rangle _{1}&=&\frac{1}{2}(\alpha'|a\rangle_{0}
+i\beta'|b\rangle_{\delta t})(\alpha|H\rangle _{S} +i\beta|H\rangle
_{L})_{1},\nonumber\\
|\psi\rangle _{2}&=&\frac{1}{2}(\alpha'|a\rangle_{0}
+i\beta'|b\rangle_{\delta t})
(i\alpha|H\rangle_{S}+\beta|H\rangle_{L})_{2}.
\end{eqnarray}
Two output ports from BS$_{2}$ are denoted by the subscripts 1 and
2, as shown in Fig. \ref{figure1}. Similar to the previous analysis
for paths A and B, we only consider channel 1 for simplicity.
Because both periods of time delay mentioned above are much less
than the phase shift, we can assume that the infection caused by
noise is the same for different parts. The change of states could be
described by a unitary operation and an overall phase  shift
\begin{eqnarray}             
|\psi\rangle&\xrightarrow[]{channel}& e^{i\theta}|\psi\rangle,
\nonumber\\
|H\rangle&\xrightarrow[]{channel}&\gamma|H\rangle +\eta|V\rangle,
\end{eqnarray}
where  $|\gamma|^2+|\eta|^2=1$. Then through  channel 1, the
evolution of the state could be written as
\begin{eqnarray}              
|\psi\rangle _{1}&=&\frac{1}{2}(\alpha'|a\rangle_{0}
+i\beta'|b\rangle_{\delta t})(\alpha|H\rangle _{S}+i\beta|H\rangle
_{L})
\nonumber \\
&\xrightarrow[]{channel}&\frac{1}{2}e^{i\theta}(\alpha'|a\rangle_{0}
+i\beta'|b\rangle_{\delta t})(\gamma\alpha|H\rangle
_{S}+\eta\alpha|V\rangle _{S} +i\gamma\beta|H\rangle
_{L}+i\eta\beta|V\rangle _{L}).
\end{eqnarray}

The decoding system, which is symmetric to the encoding system, can
transform the photon state from channel 1, which is noisy, as
follows:
\begin{eqnarray}               
|\psi'\rangle_1
&\xrightarrow[HWP]{BS_{3}}&\frac{1}{2\sqrt{2}}e^{i\theta}(\alpha'|a\rangle_{0}
+i\beta'|b\rangle_{\delta t}) (\alpha\gamma|H\rangle
_{SS}+\alpha\eta|V\rangle _{SS} +i\beta\gamma|H\rangle
_{LS}+i\beta\eta|V\rangle _{LS})
\nonumber\\
&&+\frac{1}{2\sqrt{2}}e^{i\theta}(\alpha'|a\rangle_{0}
+i\beta'|b\rangle_{\delta t}) (\alpha\gamma
i|V\rangle_{SL}+\alpha\eta i|H\rangle_{SL}
-\beta\gamma|V\rangle_{LL}-\beta\eta|H\rangle_{LL})
\nonumber\\
&\xrightarrow[]{PBS_2}&\frac{\gamma}{2\sqrt{2}}e^{i\theta}(\alpha'|a\rangle_{0}
+i\beta'|b\rangle_{\delta t})
[\alpha|H\rangle_{SS}-\beta|V\rangle_{LL}+i\underline{(\alpha|V\rangle_{SL}
+\beta|H\rangle_{LS})}]_{a'}
\nonumber\\
&&+\frac{\eta}{2\sqrt{2}}e^{i\theta}(\alpha'|a\rangle_{0}+i\beta'|b\rangle_{\delta
t}) [\alpha|V\rangle
_{SS}-\beta|H\rangle_{LL}+i\underline{(\alpha|H\rangle _{SL}
+\beta|V\rangle_{LS})}]_{b'}
\nonumber\\
&\xrightarrow[]{BS_4}&\frac{\gamma}{4}e^{i\theta}[\alpha'|a\rangle_{0}
-\beta'|b\rangle_{2\delta t}+i\underline{(\alpha'|a\rangle_{\delta
t} +\beta'|b\rangle_{\delta t})}]
[\alpha|H\rangle_{SS}-\beta|V\rangle_{LL}+i\underline{(\alpha|V\rangle_{SL}
+\beta|H\rangle_{LS})}]_{a'}
\nonumber\\
&&+\frac{\eta}{4}e^{i\theta}[\alpha'|a\rangle_{0}-\beta'|b\rangle_{2\delta
t} +i\underline{(\alpha'|a\rangle_{\delta t}+\beta'|b\rangle_{\delta
t})}] [\alpha|V\rangle
_{SS}-\beta|H\rangle_{LL}+i\underline{(\alpha|H\rangle _{SL}
+\beta|V\rangle_{LS})}]_{b'}. \label{eqn4}
\end{eqnarray}
The subscripts $a'$ and $b'$ label two outputs from the UPI system,
and the states in two paths differ by a unitary transformation.

Through the last BS, the two spatial modes are separated once more,
as displayed  in Fig. \ref{figure1}. As long as the time delay
caused by  BS$_{4}$ is equal to the initial time delay  $\delta t$,
states in the spatial mode will arrive at three definite times.
Among these times, we only consider the states in the middle time,
which carries uncorrupted information about the original quantum
state. The same method could be used in the analysis of polarization
modes, where we will focus on states in the $LS$ and $SL$ modes.
Moreover, the time delay caused by the UPI system is at least two
times larger than $\delta t$ in the case of the interference of two
DOFs. The uncorrupted states are underlined in Equation
\eqref{eqn4}. Especially, for port $a'$, a bit-flip operation on the
polarization DOF is necessary to achieve the initial polarization
state.

The success probability of obtaining the uncorrupted state for each
of the ports $a'$ and $b'$ is $\frac{1}{16}$. Notice that two main
paths A and B are separately distributed to two channels. The total
success probability is $\frac{1}{4}$.

\begin{figure*}[htb]                    
\centering
\includegraphics[width=16 cm]{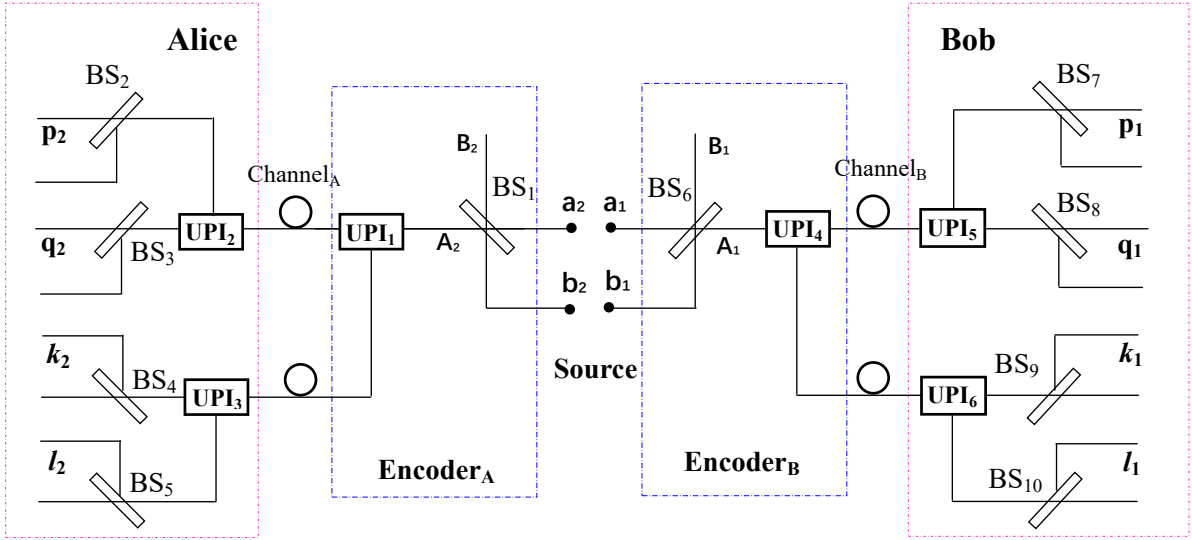}
\caption{ Schematic diagram of the error-correcting distribution of
a pair of hyperentangled photons with spatial-polarization DOFs.
$a_j$ and $b_j$ represent the two spatial modes of the photon $j$
($j=1,2$). $A_j$ and $B_j$ are two different paths for the photon
$j$. UPI is an unbalanced polarizing interferometer, as shown in
Fig. \ref{figure1}. The circles represent the channels over which
the photons are separately transmitted to the two parties Alice and
Bob. The hyperentangled photon pair $12$ in the state
$|\psi\rangle=(\alpha'|a_1a_2\rangle+\beta'|b_1b_2\rangle)_{12}(\alpha|HH\rangle+\beta|VV\rangle)_{12}$
is produced from the quantum source in the middle node and the two
photons 1 and 2 are transmitted to the two parties Bob and Alice,
respectively. Before the photons are transmitted over the
optical-fiber channels, encoders are used to code their quantum
states. After the parties Alice and Bob receive the photons, they
decode the quantum state with decoders.}\label{figure2}
\end{figure*}

\section{Error-rejecting transmission of
two entangled photons in spatial-polarization hyperentangled modes}

Now, let us consider the error-rejecting transmission of a pair of
hyperentangled photons. The schematic diagram of our scheme for the
error-correcting distribution of a pair of hyperentangled photons
with spatial-polarization DOFs is shown in Fig. \ref{figure2}. Two
entangled photons marked 1 and 2 are distributed to two identical
encoding and decoding systems and arrive at four possible ports $p$,
$q$, $k$, and $l$, respectively. Suppose that the initial state is
described by
$|\psi\rangle=(\alpha'|a_1a_2\rangle+\beta'|b_1b_2\rangle)_{12}(\alpha|HH\rangle+\beta|VV\rangle)_{12}$.

In the following discussion, we will analyze the transformation of
the states $|a\rangle$, $|b\rangle$, $|H\rangle$, and $|V\rangle$,
separately. The effect of BS$_{1}$ and BS$_{6}$ in Fig.
\ref{figure2}  can be regarded as adding a Hadamard operation to
photons in the spatial mode. The change in the states $|a\rangle$
($|a_1\rangle$ and $|a_2\rangle$) and $|b\rangle$ ($|b_1\rangle$ and
$|b_2\rangle$) could be written as
\begin{eqnarray}              
|a\rangle&\xrightarrow[]{BS}&\frac{1}{\sqrt{2}}(|a\rangle_{A}+i|a\rangle_{B})
\xrightarrow[]{channel}\frac{1}{\sqrt{2}}(e^{i\theta_{A}}|a\rangle_{A}
+ie^{i\theta_{B}}|a\rangle_{B})
\nonumber\\
&\xrightarrow[]{BS}&\frac{1}{2}(e^{i\theta_{A}}|a\rangle_{A}+ie^{i\theta_{B}}|a\rangle_{B}
+\underline{ie^{i\theta_{A}}|a\rangle_{\delta
tA}-e^{i\theta_{B}}|a\rangle_{\delta tB}}),
\nonumber\\
|b\rangle&\xrightarrow[]{BS}&\frac{1}{\sqrt{2}}(i|b\rangle_{\delta tA}+|b\rangle_{\delta tB})
\xrightarrow[]{channel}\frac{1}{\sqrt{2}}(ie^{i\theta_{A}}|b\rangle_{\delta tA}
+e^{i\theta_{B}}|b\rangle_{\delta tB})
\nonumber\\
&\xrightarrow[]{BS}&\frac{1}{2}(ie^{i\theta_{B}}|b\rangle_{2\delta
tB} -e^{i\theta_{A}}|b\rangle_{2\delta
tA}+\underline{ie^{i\theta_{A}}|b\rangle_{\delta tA}
+e^{i\theta_{B}}|b\rangle_{\delta tB}}). \label{2-1}
\end{eqnarray}
Terms with the time delay $\delta t$ are underlined because they are
the states from which we will extract information about the initial
quantum state. Applying Equation \eqref{2-1}, we can obtain an
expression of the ultimate spatial states of the hyperentangled
photons:
\begin{eqnarray}              
&&\alpha'|a_1a_2\rangle+\beta'|b_1b_2\rangle
\nonumber\\
&\xrightarrow[]{}&\frac{\alpha'}{4}(ie^{i\theta_{A_{1}}}|a_1\rangle_{\delta
tA^{1}} -e^{i\theta_{B_{1}}}|a_2\rangle_{\delta
tB^{1}})(ie^{i\theta_{A_{2}}}|a_1\rangle_{\delta tA^{2}}
-e^{i\theta_{B_{2}}}|a_2\rangle_{\delta tB^{2}})
\nonumber\\
&&+\frac{\beta'}{4}(ie^{i\theta_{A_{1}}}|b_1\rangle_{\delta tA^{1}}
+e^{i\theta_{B_{1}}}|b_2\rangle_{\delta
tB^{1}})(ie^{i\theta_{A_{2}}}|b_1\rangle_{\delta tA^{2}}
+e^{i\theta_{B_{2}}}|b_2\rangle_{\delta tB^{2}})
\nonumber\\
&=&\frac{i}{4}e^{i(\theta_{A_{1}} +\theta_{B_{2}})}(-\alpha'
|a_1\rangle_{\delta tA^{1}}|a_2\rangle_{\delta tB^{2}} +\beta'
|b_1\rangle_{\delta tA^{1}}|b_2\rangle_{\delta tB^{2}})
\nonumber\\
&&+\frac{1}{4}e^{i(\theta_{A_{1}}+\theta_{A_{2}})}(-\alpha'
|a_1\rangle_{\delta tA^{1}}|a_2\rangle_{\delta tA^{2}} -\beta'
|b_1\rangle_{\delta tA^{1}}|b_2\rangle_{\delta tA^{2}})
\nonumber\\
&&+\frac{i}{4}e^{i(\theta_{B_{1}}+\theta_{A_{2}})}(-\alpha'
|a_1\rangle_{\delta tB^{1}}|a_2\rangle_{\delta tA^{2}} +\beta'
|b_1\rangle_{\delta tB^{1}}|b_2\rangle_{\delta tA^{2}})
\nonumber\\
&&+\frac{1}{4}e^{i(\theta_{B_{1}}+\theta_{B_{2}})}(\alpha'
|a_1\rangle_{\delta tB^{1}}|a_2\rangle_{\delta tB^{2}} +\beta'
|b_1\rangle _{\delta tB^{1}}|b_2\rangle_{\delta tB^{2}}).
\label{2-2}
\end{eqnarray}
For simplicity, we neglect the states that arrive too late ($2\delta
t$) or too early.  The success probability of receiving the
uncorrupted states in spatial modes is $\frac{1}{4}$ in total, which
is clearly displayed in Equation \eqref{2-2}. For each port, a
definite unitary operation must be applied to rebuild the original
state.

The same approach could be applied to the analysis of polarization
states, which is performed as follows:
\begin{eqnarray}              
|H\rangle&\xrightarrow[]{UPI}&\frac{1}{\sqrt{2}}(|H\rangle_{S_{1}}+i|H\rangle_{S_{2}})
\nonumber\\
&\xrightarrow[]{channel}&\frac{1}{\sqrt{2}}(\gamma|H\rangle_{S}+\eta|V\rangle_{S}
+i\gamma'|H\rangle_{S}
+i\eta'|V\rangle_{S})
\nonumber\\
&\xrightarrow[]{UPI}&\frac{1}{2}[\gamma(|H\rangle_{SS}+i|V\rangle_{SL})_{p}
+\eta(|V\rangle_{SS}+|H\rangle_{SL})_{q}+i\gamma'(|H\rangle_{SS}
+|V\rangle_{SL})_{k}+i\eta'(|V\rangle_{SS}+|H\rangle_{SL})_{l}],
\nonumber\\
|V\rangle&\xrightarrow[]{UPI}&\frac{1}{\sqrt{2}}(|H\rangle_{L_{2}}+i|H\rangle_{L_{1}})
\nonumber\\
&\xrightarrow[]{channel}&\frac{1}{\sqrt{2}}(\gamma'|H\rangle_{L}+\eta'|V\rangle_{L}
+i\gamma|H\rangle_{L}+i\eta|V\rangle_{L})
\nonumber\\
&\xrightarrow[]{UPI}&\frac{1}{2}[\gamma(i|H\rangle_{LS}-|V\rangle_{LL})_{p}
+\eta(i|V\rangle_{LS}-|H\rangle_{LL})_{q}+\gamma'(|H\rangle_{LS}+i|V\rangle_{LL})_{k}
+\eta'(|V\rangle_{LS}+i|H\rangle_{LL})_{l}]. \label{2-3}
\end{eqnarray}
Notice that $\eta,\gamma$ and $\eta',\gamma'$ describe the effect of
the collective noise from different channels satisfying
$|\eta|^{2}+|\gamma|^{2}=|\eta'|^{2}+|\gamma'|^{2}=1$. Again we only
focus on those terms with the subscripts $SL$, and obtain the
uncorrupted states
\begin{eqnarray}              
&&\alpha|HH\rangle+\beta|VV\rangle
\nonumber\\
&\xrightarrow[]{}&\frac{\alpha}{4}(i\gamma_{1}|V\rangle_{p_{1}}
+\eta_{1}|H\rangle_{q_{1}}+i\gamma_{1}^{'}|V\rangle_{k_{1}}
+i\eta_{1}^{'}|H\rangle_{l_{1}})(i\gamma_{2}|V\rangle_{p_{2}}
+\eta_{2}|H\rangle_{q_{2}}+i\gamma_{2}^{'}|V\rangle_{k_{2}}+i\eta_{2}^{'}|H\rangle_{l_{2}})
\nonumber\\
&&+\frac{\beta}{4}(i\gamma_{1}|H\rangle_{p_{1}}+i\eta_{1}|V\rangle_{q_{1}}
+\gamma_{1}^{'}|H\rangle_{k_{1}}+\eta_{1}^{'}|V\rangle_{l_{1}})(i\gamma_{2}|H\rangle_{p_{2}}
+i\eta_{2}|V\rangle_{q_{2}}+\gamma_{2}^{'}|H\rangle_{k_{2}}+\eta_{2}^{'}|V\rangle_{l_{2}})
\nonumber\\
&=&-\frac{\gamma_{1}\gamma_{2}}{4}(\alpha|VV\rangle+\beta|HH\rangle)_{p_{1}p_{2}}
+\frac{\eta_{1}\eta_{2}}{4}(\alpha|HH\rangle-\beta|VV\rangle)_{q_{1}q_{2}}
+\frac{\gamma_{1}^{'}\gamma_{2}^{'}}{4}(-\alpha|VV\rangle+\beta|HH\rangle)_{k_{1}k_{2}}
\nonumber\\
&&+\frac{\eta_{1}^{'}\eta_{2}^{'}}{4}(-\alpha|HH\rangle+\beta|VV\rangle)_{l_{1}l_{2}}
+\frac{i\gamma_{1}\eta_{2}}{4}(\alpha|VH\rangle+i\beta|HV\rangle)_{p_{1}q_{2}}
+\frac{i\gamma_{1}\gamma_{2}^{'}}{4}(i\alpha|VV\rangle+\beta|HH\rangle)_{p_{1}k_{2}}
\nonumber\\
&&+\frac{i\gamma_{1}\eta_{2}^{'}}{4}(i\alpha|VH\rangle+\beta|HV\rangle)_{p_{1}l_{2}}
+\frac{i\gamma_{2}\eta_{1}}{4}(\alpha|HV\rangle+i\beta|VH\rangle)_{q_{1}p_{2}}
+\frac{i\eta_{1}\gamma_{2}^{'}}{4}(\alpha|HV\rangle+\beta|VH\rangle)_{q_{1}k_{2}}
\nonumber\\
&&+\frac{i\eta_{1}\eta_{2}^{'}}{4}(\alpha|HH\rangle+\beta|VV\rangle)_{q_{1}l_{2}}
+\frac{i\gamma_{1}^{'}\eta_{2}}{4}(\alpha|VH\rangle+\beta|HV\rangle)_{k_{1}q_{2}}
+\frac{i\gamma_{1}^{'}\gamma_{2}}{4}(i\alpha|VV\rangle+\beta|HH\rangle)_{k_{1}p_{2}}
\nonumber\\
&&+\frac{\eta_{2}^{'}\gamma_{1}^{'}}{4}(-\alpha|VH\rangle+\beta|HV\rangle)_{k_{1}l_{2}}
+\frac{i\gamma_{2}\eta_{1}^{'}}{4}(i\alpha|HV\rangle+\beta|VH\rangle)_{l_{1}p_{2}}
+\frac{i\eta_{1}^{'}\eta_{2}}{4}(\alpha|HH\rangle+\beta|VV\rangle)_{l_{1}q_{2}}
\nonumber\\
&&+\frac{\eta_{1}^{'}\gamma_{2}^{'}}{4}(-\alpha|HV\rangle+\beta|VH\rangle)_{l_{1}k_{2}}.
\label{2-4}
\end{eqnarray}

In reality, each photon includes four possible ports and a total of
$4^{2}$ combinations may be obtained. From Equation \eqref{2-4}, we
know that the success probability for polarization modes is
$\frac{1}{4}$. Taking the spatial mode into consideration, the
overall probability for obtaining a pair of uncorrupted
hyperentangled photons is $\frac{1}{16}$.

\section{Discussion and Conclusion}

Generalizing our error-rejecting transmission scheme is
straightforward in the case of hyperentangled  $n$-photon systems.
Suppose that the initial state of hyperentangled $n$ photons is
\begin{equation}              
|\psi\rangle=(\alpha'| a_{1}a_{2}\cdots a_{n}\rangle
+\beta'|b_{1}b_{2}\cdots b_{n}\rangle)(\alpha|H_{1}H_{2}\cdots
H_{n}\rangle +\beta|V_{1}V_{2}\cdots V_{n}\rangle). \label{3-1}
\end{equation}
By applying  Equation (\ref{2-1}), the evolution of the spatial
state is
\begin{eqnarray}             
|\psi\rangle_{s}&=&\alpha'| a_{1}a_{2}\cdots
a_{n}\rangle+\beta'|b_{1}b_{2}\cdots b_{n}\rangle
\nonumber \\
&\xrightarrow[]{}&\frac{\alpha'}{2^{n}}\prod \limits_{j=1}^{n} (ie^{i\theta_{A_{j}}}
|a\rangle_{\delta tA^{j}}-e^{i\theta_{B_{j}}}|a\rangle_{\delta tB^{j}})
+\frac{\beta'}{2^{n}}\prod\limits_{j=1}^{n}(ie^{i\theta_{A_{j}}}|b\rangle_{\delta tA^{j}}
+e^{i\theta_{B_{j}}}|b\rangle_{\delta tB^{j}})
\nonumber\\
&=&\sum\limits_{m=0}^n \ \sum\limits_{\forall a_{1},\cdots, a_{m}}\frac{e^{i\Theta}}{2^n}
(\alpha'(-1)^{n-m}i^m|a\rangle_{A^1}|a\rangle_{A^2}\cdots|a\rangle_{A^m}|a\rangle_{B^1}
\cdots|a\rangle_{B^{n-m}}
\nonumber\\
&&\hspace{25mm}+\beta'i^m|b\rangle_{A^1}|b\rangle_{A^2}\cdots|b\rangle_{A^m}|b\rangle_{b^1}
\cdots|b\rangle_{b^{n-m}})
\label{3-2}
\end{eqnarray}
Here the symbol $\sum\limits_{\forall a_{1},\cdots, a_{m}}$ means
the sum of all the occasions when m photons arrive at path A while
n-m photons arrive at path B. For simplicity, we replace various
phase shifts by a uniform term $e^{i\Theta}$. For each photon,  a
consistent one-to-one match between the states $|a\rangle$ and
$|b\rangle$  may be noted. Thus, a total of
$\sum_{m=0}^n\binom{n}{m}=2^n$ combinations is achieved, which
offers a success probability of  $\frac{1}{2^n}$ . For the
polarization mode, using Equation \eqref{2-3}, we get
\begin{eqnarray}             
|\psi\rangle_{p}&=&\alpha|H_{1}H_{2}\cdots
H_{n}\rangle+\beta|V_{1}V_{2}\cdots V_{n}\rangle
\nonumber\\
&\xrightarrow[]{}&\frac{\alpha'}{2^n}\prod \limits_{j=1}^{n}(i\gamma_{j}|V\rangle_{p^j}
+\eta_{j}|H\rangle_{q^j}+\gamma'_{j}i|V\rangle_{k^j}+i\eta'_{j}|H\rangle_{l^j})
+\frac{\beta'}{2^n}\prod \limits_{j=1}^{n}(i\gamma_{j}|H\rangle_{p^j}
+i\eta_{j}|V\rangle_{q^j}+\gamma'_{j}|H\rangle_{k^j}+\eta'_{j}|V\rangle_{l^j})
\nonumber\\
&=&\sum\limits_{a+b+c+d=n} \frac{\gamma\eta}{2^n}(\alpha i^{a+c+d}|V\rangle_{p^1}
\cdots|V\rangle_{p^a}|H\rangle_{q^1}\cdots|H\rangle_{q^b}|V\rangle_{k^1}
\cdots|V\rangle_{k^c}|H\rangle_{l^1}\cdots|H\rangle_{l^d}
\nonumber\\
&&\hspace{23mm}+\beta
i^{a+b}|H\rangle_{p^1}\cdots|H\rangle_{p^a}|V\rangle_{q^1}
\cdots|V\rangle_{q^b}|H\rangle_{k^1}\cdots|H\rangle_{k^c}|V\rangle_{l^1}\cdots|V\rangle_{l^d}).
\end{eqnarray}
Notice that $\gamma$ and $\eta$ represent the collective noise on
different channels,  the total success probability for transmitting
n photons in spatial-polarization hyperentangled mode is
$\frac{1}{4^n}$.


Hyperentanglement presents good applicability in high-capacity
quantum communication. The proposed self-error-rejecting photonic
qubit transmission scheme can actively reject the collective noise
in optical channels. Similar to the case where hyperentanglement
purification \cite{HEPP1,HEPP2,HEPPWangGY,DuFFHEPP,HEPP5} and
hyperentanglement concentration
\cite{HECP1,HECP2,HyperC3,HyperC4,HyperC5,HECP6,HyperC7,HyperC8,HyperC9,HECPadddu2}
are more complicated than entanglement purification
\cite{EPP1,EPP2,EPPsimon,EPPsheng1,EPPelectron,EPPhybid,EPPatomensemble,eppshengc,eppshengb}
and entanglement concentration
\cite{ECP1,ECP3a,ECP3b,ECP5,ECP6,DengECP,ECP7,ECPadddu1,EPPatom,ECPshengadd1,ECPshengadd2,ECPshengadd3,ECPad1,ECPad2,ECPad3,ECPad4,ECPad5}
for photon systems in only one DOF, this self-error-rejecting scheme
for hyperentangled photon systems is more complicated than the
self-error-rejecting scheme for photon systems in only the
polarization DOF \cite{r5}. However, it is an active method that
allows hyperentangled photon systems to overcome the collective
noise in channels. Combining hyperentanglement purification and
hyperentanglement concentration
\cite{HEPP1,HEPP2,HEPPWangGY,DuFFHEPP,HEPP5,HECP1,HECP2,HyperC3,HyperC4,HyperC5,HECP6,HyperC7,HyperC8,HyperC9,HECPadddu2},
it may be very useful for high-capacity quantum communication as
this scheme works with linear-optical elements and can, in
principle, be implemented easily with current experimental
techniques.

In our self-error-rejecting scheme for photon systems in two DOFs,
we assume the efficiency of the linear-optical elements, including
PBSs, BSs, and HWPs, to be perfect, i.e., no photon loss occurs in
these linear-optical elements. In practical applications of this
self-error-rejecting scheme, the elements do not work in ideal
conditions, and photon loss is one of the main challenges for its
application. Another challenge that must be addressed is the
accurate time delay of all over relatively long period of times,
which will affect the success probability of the interference
between two wave packets in decoders. In principle, these two main
challenges can be overcome with optical chips, which can integrate
linear-optical elements to control the delay time accurately and
depress photon loss.

In summary, we have proposed a self-error-rejecting qubit
transmission scheme for an N-photon system in both the polarization
and spatial states over noisy channels using linear-optical elements
only. The polarization and spatial states are easily influenced by
the collective noise. By converting these states into different time
bins, the influence of noise could be transformed into a global
phase, thereby enabling easy selection of uncorrupted states during
the detection process. Moreover, our scheme promises a stable
success probability without suffering from the influence of the
coefficients of noisy channels. This scheme may be very useful in
practical high-capacity quantum communication with photons in two or
more DOFs.

\section*{ACKNOWLEDGMENTS}

This work is supported by the National Natural Science Foundation of
China under Grants No. 61675028 and No. 11674033, the Fundamental
Research Funds for the Central Universities under Grant No.
2015KJJCA01, and the National High Technology Research and
Development Program of China under Grant No. 2013AA122902.


\begin{thebibliography}{99}

\bibitem{QT1} C. H. Bennett,  G. Brassard, C. Cr\'{e}peau, R. Jozsa, A. Peres, and W. K.
 Wootters,  Phys Rev Lett. \textbf{70}, 1895 (1993).


\bibitem{QT2} Q. Ai,  Sci  Bull. \textbf{61}, 110 (2016).


\bibitem{DC1} C. H.  Bennett  and  S. J. Wiesner,  Phys Rev Lett. \textbf{69}, 2881
(1992).


\bibitem{DC2} X. S. Liu,  G. L. Long,  D. M. Tong,  and  F. Li,  Phys Rev A. \textbf{65},
022304 (2002).


\bibitem{QKD0} C. H. Bennett  and  G. Brassard,   Proceedings of IEEE International
Conference on Computers, Systems and Signal Processing (Bangalore,
1984) P.175.


\bibitem{QKD1} A. K. Ekert, Phys Rev Lett. \textbf{67}, 661 (1991).


\bibitem{QKD2} C. H. Bennett, G. Brassard, and  N. D. Mermin, Phys Rev A.  \textbf{68}, 557 (1992).


\bibitem{QKD3} X. H. Li, F. G. Deng, and H. Y. Zhou, Phys Rev A.  \textbf{78}, 022321 (2008).


\bibitem{QSS1} M. Hillery, V. Bu\v{z}ek, and A. Berthiaume, Phys Rev A. \textbf{59}, 1829 (1999).


\bibitem{QSDC1} G. L. Long and X. S. Liu, Phys Rev A. \textbf{65}, 032302 (2002).


\bibitem{QSDC2} F. G. Deng, G. L. Long, and X. S. Liu,  Phys Rev A. \textbf{68}, 042317 (2003).


\bibitem{QSDC3} F. G. Deng and G. L. Long,  Phys Rev A. \textbf{69}, 052319 (2004).


\bibitem{QSDC4} C. Wang, F. G. Deng, Y. S. Li, X. S. Liu, and G. L. Long,
Phys Rev A. \textbf{71}, 044305 (2005).


\bibitem{QSDC5} J. Y. Hu, B. Yu, M. Y. Jing, L. T. Xiao, S. T. Jia, G. Q. Qin, and
 G. L. Long,   Light: Sci Appl. \textbf{5}, e16144 (2016).


\bibitem{QSDC6} W. Zhang, D. S. Ding, Y. B. Sheng, L. Zhou, B. S. Shi, and G. C. Guo,
Phys Rev Lett. \textbf{118}, 220501 (2017).


\bibitem{QSDCreview} X. H. Li, Acta Phys Sin. \textbf{64}, 160307 (2015).



\bibitem{r2} D. Kalamidas, Phys Lett A. \textbf{343}, 331 (2005).


\bibitem{r3} T. Yamomoto, J. Shimamura, S. K. Ozdemir, M. Koashi, and N. Imoto,
Phys Rev Lett. \textbf{95}, 040503 (2005).



\bibitem{r5} X. H. Li, F. G. Deng, and H. Y. Zhou,
Appl Phys Lett. \textbf{91}, 1444101 (2007).


\bibitem{dengreview} F. G. Deng, B. C. Ren, and X. H. Li, Sci Bullet. \textbf{62}, 46 (2017).


\bibitem{prep1} J. T.  Barreiro, N. K. Langford, N. A. Peters, and P. G. Kwiat,
Phys Rev Lett. \textbf{95}, 260501 (2005).


\bibitem{prep2} M. Barbieri, C. Cinelli, P. Mataloni, F. De Martini,
Phys Rev A. \textbf{72}, 052110 (2005).



\bibitem{prep3} R. Ceccarelli, G. Vallone, F. De Martini, P. Mataloni, and
 A. Cabello,  Phys Rev Lett. \textbf{103}, 160401 (2009).


\bibitem{prep4} G. Vallone, R. Ceccarelli, F. De Martini, and P. Mataloni,
Phys Rev A. \textbf{79}, 030301(R) (2009).


\bibitem{prep5} A. Rossi, G. Vallone, A. Chiuri, F. De Martini, and P. Mataloni,
Phys Rev Lett. \textbf{102}, 153902 (2009).


\bibitem{prep6} D. Bhatti, J. von Zanthier, and G. S. Agarwal,
Phys Rev A. \textbf{91}, 062303 (2015).





\bibitem{HBSA1} Y. B. Sheng, F. G. Deng, and G. L. Long, Phys
Rev A. \textbf{82}, 032318 (2010).


\bibitem{HBSA2} B. C. Ren, H. R. Wei, M. Hua, T. Li, and F. G. Deng,
Opt Express. \textbf{20}, 24664 (2012).


\bibitem{HBSA3} T. J. Wang, Y. Lu, and G. L. Long,   Phys Rev A. \textbf{86}, 042337 (2012).

\bibitem{HBSA4} Q. Liu and M. Zhang,  Phys Rev A. \textbf{91}, 062321 (2015).




\bibitem{HBSA6} X. H. Li and S. Ghose,  Phys Rev A. \textbf{93}, 022302 (2016).


\bibitem{HBSA7}  G. Y. Wang, Q. Ai, B. C. Ren, T. Li, and
 F. G. Deng, Opt Express. \textbf{24}, 28444 (2016).







\bibitem{hypercnot1} B. C. Ren and F G. Deng, Sci Rep. \textbf{4}, 4623 (2014).


\bibitem{hypercnot2} B. C. Ren, G. Y. Wang, and F. G. Deng,  Phys Rev A. \textbf{91}, 032328 (2015).


\bibitem{hypercnot3}
 T. Li  and G. L. Long, Phys Rev A. \textbf{94}, 022343 (2016).


\bibitem{hypercnot4}
B. C. Ren and F. G. Deng,  Opt Express. \textbf{25}, 10863 (2017).



\bibitem{HESC} J. T. Barreiro, T. C. Wei, and P. G. Kwiat, Nat Phys. \textbf{4}, 282
(2008).

\bibitem{DEPP1} Y. B. Sheng and F. G. Deng, Phys Rev A. \textbf{81}, 032307 (2010).

\bibitem{DEPP2} Y. B. Sheng and F. G. Deng,  Phys Rev A. \textbf{82}, 044305 (2010).

\bibitem{DEPP3} X. H. Li,   Phys Rev A. \textbf{82}, 044304 (2010).

\bibitem{DEPP4} F. G. Deng,   Phys Rev A. \textbf{83}, 062316 (2011).

\bibitem{EPPblind} Y. B. Sheng and  L. Zhou,  Sci Rep. \textbf{5}, 7815 (2015).


\bibitem{wangrepeater} T. J. Wang, S. Y. Song, and G. L. Long,   Phys Rev A. \textbf{85}, 062311
(2012).

\bibitem{HEPP1} B. C. Ren and F. G. Deng,   Laser Phys Lett. \textbf{10}, 115201
(2013).

\bibitem{HEPP2} B. C. Ren, F. F. Du, and F. G. Deng,  Phys Rev A. \textbf{90},
052309 (2014).

\bibitem{HEPPWangGY} G. Y. Wang, Q. Liu, and F. G. Deng,   Phys Rev A. \textbf{94},
032319 (2016).

\bibitem{DuFFHEPP} F. F. Du, T. Li, and G. L. Long, Ann Phys. \textbf{375}, 105  (2016).

\bibitem{HEPP5} T. J. Wang, L. L. Liu, R. Zhang, C. Cao, and C. Wang,
Opt Express. \textbf{23}, 9284 (2015).

\bibitem{HECP1} B. C. Ren, F. F. Du, and F. G. Deng,   Phys Rev A. \textbf{88},
012302 (2013).

\bibitem{HECP2} B. C. Ren  and  G. L. Long,   Opt Express. \textbf{22}, 6547 (2014).

\bibitem{HyperC3} X. H. Li  and  S. Ghose,   Laser Phys Lett. \textbf{11}, 125201
(2014).

\bibitem{HyperC4} X. H. Li  and  S. Ghose,  Opt Express. \textbf{23}, 3550  (2015).

\bibitem{HyperC5} X. H. Li  and  S. Ghose,   Phys Rev A. \textbf{91},
062302 (2015).

\bibitem{HECP6} B. C. Ren  and  G. L. Long,   Sci Rep.  \textbf{5}, 16444 (2015).

\bibitem{HyperC7}
 C. Cao, T. J. Wang, S. C. Mi, R. Zhang, and C. Wang,   Ann Phys. \textbf{369}, 128 (2016).

\bibitem{HyperC8} L. L.
Fan, Y. Xia, and J. Song, Quantum Inf Process. \textbf{13}, 1967
(2014).

\bibitem{HyperC9} H. J. Liu, Y. Xia, and  J. Song,
Quantum Inf Process. \textbf{15}, 2033 (2016).


\bibitem{HECPadddu2}  F. F. Du, F. G. Deng,  and G. L. Long, Sci Rep.
\textbf{6}, 35922 (2016).



\bibitem{EPP1}  C. H.
Bennett,  G. Brassard, S. Popescu, B. Schumacher, J. A. Smolin, and
 W. K. Wootters,  Phys Rev Lett.  \textbf{76}, 722 (1996).



\bibitem{EPP2} J. W. Pan, C. Simon, \v{C}. Brukner, and A. Zeilinger,
Nature. \textbf{410}, 1067 (2001).

\bibitem{EPPsimon} C. Simon  and J. W. Pan, Phys Rev Lett.  \textbf{89}, 257901 (2002).

\bibitem{EPPsheng1}
Y. B. Sheng, F. G. Deng, and H. Y. Zhou,  Phys Rev A. \textbf{77},
042308 (2008).

\bibitem{EPPelectron} C. Wang,  Y. Zhang, and G. S. Jin, Phys Rev A. \textbf{84}, 032307 (2011).

\bibitem{EPPhybid} Y. B. Sheng, L. Zhou,  G. L. Long,  Phys Rev A. \textbf{88}, 022302
(2013).

\bibitem{EPPatomensemble} T. Li, G. J. Yang, and F. G. Deng,   Opt Express.
 \textbf{22}, 23897 (2014).

\bibitem{eppshengc}  L. Zhou  and  Y. B. Sheng, Laser Phys Lett. \textbf{12}, 045203 (2015).

\bibitem{eppshengb}  L. Zhou  and  Y. B. Sheng,
Sci Rep. \textbf{6}, 28813  (2016).



\bibitem{ECP1}  C. H. Bennett,  H. J. Bernstein, S. Popescu,  and B. Schumacher,  Phys Rev
A. \textbf{53}, 2046 (1996).



\bibitem{ECP3a}  Z. Zhao, J. W. Pan, and M. S. Zhan, Phys Rev A.
\textbf{64}, 014301 (2001).



\bibitem{ECP3b} T. Yamamoto, M. Koashi, and N. Imoto, Phys Rev A. \textbf{64},  012304
(2001).


\bibitem{ECP5}  Y. B. Sheng,  F. G. Deng, and H. Y. Zhou, Phys Rev A.  \textbf{77}, 062325
(2008).



\bibitem{ECP6} Y. B. Sheng,  L. Zhou, S. M. Zhao, and B. Y. Zheng, Phys Rev A.
 \textbf{85}, 012307 (2012).


\bibitem{DengECP}  F. G. Deng, Phys Rev A.  \textbf{85},  022311 (2012).


\bibitem{ECP7} Y. B. Sheng,  L. Zhou, and S. M. Zhao, Phys Rev A. \textbf{85}, 042302 (2012).


\bibitem{ECPadddu1} F. F. Du and F. G. Deng,  Sci  China-Phys  Mech  Astron.
 \textbf{58}, 040303 (2015).


\bibitem{EPPatom}   C. Cao,  C. Wang, L. Y. He, and R. Zhang, Opt Express. \textbf{21}, 4093 (2013).



\bibitem{ECPshengadd1}  C. Wang, W. W. Shen, S. C. Mi, Y. Zhang, and T. J.
Wang,  Sci  Bull.  \textbf{60}, 2016 (2015).


\bibitem{ECPshengadd2} C. Cao, X. Chen, Y. W. Duan, L. Fan,  R. Zhang,  T. J. Wang,
  C. Wang,   Sci  China-Phys  Mech  Astron.  \textbf{59}, 100315 (2016).


\bibitem{ECPshengadd3} Y. B. Sheng, J. Pan, R. Guo,  L. Zhou, and L.
Wang, Sci  China-Phys  Mech  Astron.  \textbf{58}, 060301 (2015).


\bibitem{ECPad1} F. F. Du and G. L. Long,  Quantum Inf Process. \textbf{16}, 26 (2017).



\bibitem{ECPad2} H. J. Liu, L. L. Fan, Y. Xia, and J. Song,
Quantum Inf Process. \textbf{14}, 2909 (2015).


\bibitem{ECPad3} A. Banerjee, C. Shukla,  A. Pathak,   Quantum Inf
Process. \textbf{14}, 4523 (2015).


\bibitem{ECPad4} C. Shukla, A. Banerjee,  A. Pathak,
Quantum Inf Process. \textbf{14}, 2077 (2015).


\bibitem{ECPad5} J. Pan,  L. Zhou, S. P. Gu, X. F. Wang, Y. B. Sheng, and Q.
Wang, Quantum Inf Process. \textbf{15}, 1669 (2016).








\end{thebibliography}
\end{document}